\documentstyle[12pt,psfig]{article}
\begin{document}
\setlength{\unitlength}{1mm}

{\hfill   August 1998 }

{\hfill    Alberta Thy 13-98}

{\hfill    hep-th/9808132} \vspace*{2cm} \\
\begin{center}
{\Large\bf Planckian $AdS_2 \times S_2$ space is an exact
solution of the semiclassical Einstein equations}
\end{center}
\begin{center} 
Sergey N.~Solodukhin\footnote{
Present address: Spinoza Institute,
University of Utrecht, 
Leuvenlaan 4, P.O.Box 80.195,
3508 TD Utrecht,
the Netherlands}
\end{center}
\begin{center}
{\it Department of Physics, University of Alberta, Edmonton, Alberta, T6G 2J1, 
Canada}  \\
e-mail: sergey@phys.ualberta.ca
\end{center}
\vspace*{1cm}
\begin{abstract}
\noindent
The  product space configuration $AdS_2\times S_2$ 
(with  $l$ and $r$ being radiuses  of the components)
 carrying the electric charge $Q$
is demonstrated to be an exact solution of the semiclassical
Einstein equations in presence of the Maxwell field.
If the logarithmic  UV divergences are absent
in the four-dimensional theory the solution  we find is  identical to the classical Bertotti-Robinson
space ($r=l=Q$) with no quantum corrections added.
In general, the analysis involves the quadratic curvature coupling $\lambda$ appearing
in the effective action. The solutions  we find are of the following
types:
i) (for arbitrary $\lambda$) charged configuration which is  quantum
deformation of the Bertotti-Robinson space; ii) ($\lambda >\lambda_{cr}$)
$Q=0$ configuration with $l$ and $r$ being of the Planck order; iii)
($\lambda<\lambda_{cr}$) $Q\neq 0$ configuration ($l$ and $r$ are
of the Planck order)  not connected analytically to the
Bertotti-Robinson space. The interpretation of the solutions  obtained
and  an
indication on the internal structure
of the Schwarzschild black hole are discussed.
\end{abstract}
\newpage
\baselineskip=.8cm
It is an interesting and important problem as how the known
solutions of classical Einstein equations are modified in 
the quantum domain. It is believed that the quantum corrections
become essential when the space-time curvature is of the Planck order
and they may change drastically the space-time structure
appearing in the classical theory \cite{FMM}, \cite{Frol-Vilk}, 
\cite{KS}. Particularly, 
the black hole geometry may be corrected at the Planck distances in
the way that the singularity at $r=0$ is 
replaced by a regular manifold. On the other hand, the quantum corrections may occur to be important on the last stage
of the gravitational collapse and be highly
influential on the ultimate fate of the black hole.
However, the current status of the theory
does not allow us to make any definite
conclusion about realization of these expectations. The main 
problem is that we still do not have the consistent theory of quantum gravity in the framework of which all these questions might be answered
consistently. The best  we can do is to develop
the semiclassical approximation when the gravitational field (metric
$g_{\mu\nu}$ on space-time) is still treated as a classical object
while the matter fields are quantized on its background. In this case
configurations of the  gravitational field are governed by the ``quantum''
gravitational action
\begin{equation}
W_Q[g_{\mu\nu}]=W_{cl}+\Gamma [g_{\mu\nu}]~~,
\label{1}
\end{equation}
where $W_{cl}$ is the classical action. For the gravity coupled with 
the Maxwell field we have
\begin{equation}
W_{cl}=-{1\over 16\pi G}\left(\int_{M^4}R_{(4)}+2\int_{\partial M^4}
K_{(4)}+\int_{M^4} F^2_{\mu\nu}\right)~~,
\label{2}
\end{equation}
on the $4$-dimensional space $M^4$, $R_{(4)}$ is the four-dimensional
scalar curvature and $K_{(4)}$ is the trace of the extrinsic curvature of the boundary $\partial M^4$.

The term $\Gamma [g_{\mu\nu}]$ in (\ref{1}) is due to 
the quantum matter fields. It is highly non-local functional of $g_{\mu\nu}$
and its form is not known in general. This fact makes the analysis of the
semiclassical Einstein equations 
\begin{equation}
{\delta W_Q\over \delta g_{\mu\nu}}=0
\label{3}
\end{equation}
possible only in some approximation \cite{Y}. Note, that the functional 
$\Gamma [g_{\mu\nu}]$ is better studied in two dimensions
that makes the study of two-dimensional models \cite{2d}
so attractive.
The efficiency, however, of the approximate methods and toy models
would be considerably supported by finding at least one configuration which
is a guaranteed exact solution of the equations (\ref{3}).
In this note we find such a solution. It is  the direct 
product of two-dimensional Anti-de Sitter space $AdS_2$ and
two-dimensional sphere $S_2$ with  respectively
radius $l$ and $r$  of each component.

An indirect indication of existence of this solution comes from the analysis
of $2d$  quantum models. It was found that there always appears
a solution of the quantum gravitational equations describing
$2d$ space with constant curvature and constant value of the
dilaton field. First this fact was observed in \cite{1} for the RST model
and later found to be a feature of more general class of $2d$ models
\cite{2}. Translating this into the $4$-dimensional language the dilaton
field should be identified with the radius $r$ of the spheri-symmetric
metric
\begin{equation}
ds^2=\gamma_{\alpha\beta}(z)dz^\alpha dz^\beta+r^2(z)(d\theta^2+\sin^2\theta
d\phi^2)
\label{4}
\end{equation}
and the corresponding four-dimensional configuration would be
$AdS_2\times S_2$. In what follows we give proof both in terms of the $2d$
model and the four-dimensional theory (\ref{1})
that the space $AdS_2\times S_2$ is indeed an exact 
solution of the semiclassical gravitational equations.

Before proceeding we pause for a few remarks. The space 
$AdS_2\times S_2$ with 
\begin{equation}
l=r=Q~~,
\label{*}
\end{equation}
where $Q$ is the electric charge,
is so-called the Bertotti-Robinson space. It is known 
to be a solution of the classical Einstein equations in 
presence of the Maxwell field and can be viewed as  near-horizon
geometry in the extreme limit ($M\rightarrow Q$) of the charged
black hole metric. For $Q\neq 0$ the solution we find below 
is the quantum deformation 
of this classical solution. Remarkably, (and it is  one of our main points)
the space  of this kind remains a solution of eq.(\ref{3}) even for $Q=0$.
The radiuses $l$ and $r$ then are of the Planck order.
Actually, an  $AdS_2\times S_2$ space  (with arbitrary $r$ and $l$)
appears to describe universally the extreme limit \cite{Zas}
of a black hole configuration
both in the classical and semiclassical
theories. The  quantum black hole entropy then takes an universal form 
\cite{MS}  (dependent on $r$ and $l$)  when  the extreme limit
is approached.
The product space $AdS_2\times S_2$ has  also become recently a subject of intensive
study from different point of view in \cite{3}. It would be interesting
to exploit our results in the context of the study made 
in \cite{3}.

In our study of the equations (\ref{3}) 
we are interested in a solution from the class of the spheri-symmetric
metrics (\ref{4}). Any such metric is completely determined
by  fixing the two-dimensional metric $\gamma_{\alpha\beta}$ and
the ``dilaton'' field $r(z)$. Being considered on the class of the
four-dimensional metrics  (\ref{4}) the classical action (\ref{2}) 
takes the form of the two-dimensional theory of gravity\footnote{
We use the Euclidean signature of the metric.}  \cite{KS}, \cite{2d}
\begin{equation}
W_{cl}[\gamma_{\alpha\beta},r]=-{1\over 4G}\int_{M^2}\left(
r^2R+2(\nabla r)^2+2U(r)\right)~~,
\label{5}
\end{equation}
where $M^2$ is $2d$ space with coordinates $\{z^\alpha \}$, $R$ is $2d$
scalar curvature and $U(r)$ is the dilaton potential. When
electric charge is zero we have $U(r)=1$. Otherwise, it is $U(r)=1-{Q^2
\over r^2}$. It is easy to see that the configuration (\ref{*}) is a solution
of field equations obtained from the action (\ref{5}). Indeed, variation with respect
to dilaton $r$ (for a configuration with constant $R$ and $r$)
gives $R=-r^{-1}U'(r)=-2{Q^2\over r^4}$ while the variation with respect to metric results in vanishing the potential $U(r)$. Altogether, both conditions lead to (\ref{*}).

Now we have to find the form of the effective action $\Gamma[g_{\mu\nu}]$
in (\ref{1}). Considered on the class of spheri-symmetric metrics (\ref{4})
it becomes a functional $\Gamma[R,r, \nabla R, \nabla r]$ of two-dimensional
scalar curvature $R$, dilaton $r$ and their derivatives $\nabla R$,
$\nabla r$. Just for the illustration we first consider the functional
$\Gamma$ arising in two-dimensional case when the $2d$ massless
fields couple to the dilaton $r$.  In two dimensions, when $r$ and $R$ are constant $\Gamma$
changes as $\Gamma\rightarrow \Gamma+c\int_{M^2} R~\ln \alpha$
under the scaling transformation $r\rightarrow r\alpha$. Therefore,
the form of $\Gamma$ in two dimensions is the following
(see \cite{KS} and \cite{?})
\begin{equation}  
\Gamma [\gamma_{\alpha \beta},r]=A\int_{M^2}R\Box^{-1}R+B\int_{M^2}R\ln r
+w[\nabla r, \nabla R]~~,
\label{6}
\end{equation}
where $A$ and $B$ are constants, $w[\nabla r, \nabla R]$ is the functional
which vanishes when $r$ and $R$ are constant. It is important for the analysis
we are carrying on that variation of the term $w[\nabla r, \nabla R]$
with respect to metric $\gamma_{\alpha\beta}$ or dilaton $r$
vanishes for a configuration with constant $R$ and $r$. Therefore,
we may ignore the term $w[\nabla r, \nabla R]$
when looking for a solution  of the semiclassical equations that
describes constant curvature $2d$ space-time with constant value of
the dilaton field $r$. Variation of $W_Q[\gamma_{\alpha \beta},r]$
(\ref{1})
then gives rise to the equations
\begin{eqnarray}
&&rR+U'(r)={2GB\over r}R \nonumber \\
&&U(r)=-4AGR~~,
\label{7}
\end{eqnarray}
where $U'(r)=\partial_rU(r)$. The analysis of  the eqs.(\ref{7}) goes for
arbitrary $A$ and $B$. For simplicity, however,  we assume that $B=4A$.
Then, defying the radius $l$ of the $2d$ space as $R=-{2\over l^2}$ we
find from (\ref{7}) that
\begin{equation}
r^2=l^2=Q^2+2Al^2_{pl}~~
\label{8}
\end{equation}
where the Planck length $l_{pl}$ is defined as $l^2_{pl}=4G$. 
The solution (\ref{8}) is a quantum deformation (governed by the
Planck length $l_{pl}$) of the classical solution (\ref{*})
describing the near horizon geometry of the extreme limit of the charged black
hole. However, the solution (\ref{8}) has a new feature absent in the classical case. Namely, it is still valid even if the electric 
charge vanishes ($Q=0$). The four-dimensional space then is $AdS_2\times S_2$ with each component having the Planck order radius $\sqrt{2A}l_{pl}$. It is worth noting that the coupling to dilaton is
important for  the existence of the solution of 
the equations (\ref{7}) with $Q=0$. Otherwise,
if $B=0$  (as it is in the $2d$ model considered in \cite{2d})
in (\ref{6})  one finds from (\ref{7}) that
\begin{equation}
r^2=lQ~,~~l^2-r^2=8AG=2Al^2_{pl}~~.
\label{9}
\end{equation}
In the limit $Q\rightarrow 0$ it describes a singular (in the four-dimensional
picture) configuration
\begin{equation}
r=0~,~~l=\sqrt{2A}l_{pl}~~
\label{10}
\end{equation}
with vanishing size of the spheric component $S_2$.

\bigskip

We now want to extend this analysis to the four-dimensional case. This requires the knowledge of the structure of the four-dimensional effective action
$\Gamma [r,R,\nabla r, \nabla R]$ which as we  have already mentioned is very
complicated and not known in general. For our purposes, however,
we need to know only how $\Gamma$ depends on $r$ and $R$ ignoring
gradients $\nabla r$ and $\nabla R$. In general, the functional $\Gamma$ can be represented
in the form 
\begin{equation}
\Gamma[r,R,\nabla r, \nabla R]=\Gamma_0[r,R]+w[\nabla r, \nabla R]
\label{11}
\end{equation}
where $\Gamma_0[r,R]$ is a functional of $r$ and $R$ but not their derivatives
and $w[\nabla r, \nabla R]$ is the functional vanishing when $\nabla r=\nabla R=0$, it can be expended in powers of $\nabla r$ and $\nabla R$.
As in the $2d$ case considered above we may ignore variation of $w[\nabla r, \nabla R]$ when looking for a solution 
with constant $r$ and $R$. Thus, only variation of $\Gamma_0[r,R]$
produces the essential for our purposes part of the semiclassical Einstein equations.
The structure of $\Gamma_0[r,R]$ can be obtained by quantizing the
matter fields on the background space $AdS_2\times S_2$ with arbitrary
radius $l$ ($r$) of the component $AdS_2$ ($S_2$).
This can be carried out by, for example, the ${\it zeta}$-function method
$$
\Gamma_0=-{1\over 2}(\zeta '(0)-\zeta (0)\ln \mu^2)~~,
$$
where $\mu$ is an arbitrary  length scale and $\zeta$-function
on the product space $AdS_2\times S_2$ is defined as
\begin{equation}
\zeta (s)={1\over \Gamma (s)}\int_0^{+\infty}dtt^{s-1}Tr K_{H_2}~TrK_{S_2}~~,
\label{zeta}
\end{equation}
where $K_{H_2}(t)$ ($K_{S_2}(t)$) is the heat kernel
on $AdS_2$  ($S_2$) space. 

For concreteness we consider a four-dimensional scalar field with zero mass.
The heat kernels on spaces $AdS_2$ and $S_2$ 
are known explicitly, $Tr K_{H_2}={V_{H_2}\over 4\pi l^2}k({t\over l^2})$,
$Tr K_{S_2}=\Theta ({t\over r^2})$, and the corresponding
expressions for $k({t\over l^2})$ and $\Theta ({t\over r^2})$ can be found in
\cite{Camporesi}.
It is important  that the heat kernels on sphere 
and on Anti-de Sitter space are related by an 
analytical continuation $l^2\rightarrow -r^2$.
In particular, it  is manifested in the small
$t$ expansion
\begin{eqnarray}
&&k({t\over l^2})\simeq {l^2\over  t}(1-{t\over 3l^2}+{t^2\over 15l^4})
\nonumber \\
&&\Theta ({t\over r^2})\simeq {r^2\over t}(1+{t\over 3r^2}+{t^2\over 15r^4})~~
\label{expansion}
\end{eqnarray}
of the heat kernels.

In our case (\ref{zeta}) is function $\zeta (s, r^2,l^2)$ of the parameter
$s$ and the radiuses $r$ and $l$ of the space $S_2$ and $AdS_2$
respectively. By means of the 
known scaling property $\zeta (s, r^2, l^2)=r^{2s}\zeta (s, 1, {l^2\over r^2})$
we find that
\begin{equation}
\zeta '(0,r^2,l^2)=\zeta '(0,1,{l^2\over r^2})+\zeta (0,1,{l^2\over r^2})
\ln r^2~~.
\label{prop}
\end{equation}
The effective action  then takes the form
\begin{equation}
\Gamma_0[r,l]=-{V_{H_2}\over 2l^2}\left(C({r^2\over l^2})\ln {r^2\over G}+\lambda (\mu )C({r^2\over l^2})+
F({r^2\over l^2})\right)~~,
\label{12}
\end{equation}
where $V_{H_2}$ is volume of $AdS_2$ space, note that $V_{H_2}=l^2v$
where $v$ is the divergent dimensionless quantity. 
We denoted $F({r^2\over l^2}){V_{H_2}\over l^2}=\zeta'(0,1,{l^2\over r^2})$
and $C({r^2\over l^2}){V_{H_2}\over l^2}=\zeta (0,1,{l^2\over r^2})$
and introduced the $\mu$-dependent constant $\lambda (\mu )=\ln {G\over \mu^2}$
when derived (\ref{12}). It is known that $\zeta (0)=a_2$, where $a_2$ is a coefficient in the small $t$ expansion $Tr K_{M^4}(t)=\sum_{n=0}^\infty a_{n\over 2}
t^{(n-4)/2}$ of the heat kernel.  The coefficient $a_2$ is responsible for the
logarithmic $UV$ divergences of the effective action. Note, that the $\zeta$-function method provides us with already regularized
expression for the effective action.
The function $C(x)$
in (\ref{12}), thus,  represents the local  quadratic in curvature 
terms 
with $\lambda (\mu )$ being the corresponding
renormalized coupling in the four-dimensional effective action.
On the other hand,  the function $F(x)$
represents the non-local part of the action. 
For the massless  scalar field
we find using the expansion (\ref{expansion}) that
\begin{equation}
C(x)={1\over 60\pi}(x+{1\over x}-{5\over 3})~~,
\label{*3}
\end{equation}
where $x={r^2\over l^2}$. Note, that  the function
$C(x)$ is positive 
for $x>0$ and has minimum at $x=1$ ($C'(1)=0$).

Knowing the explicit form of the heat kernels entering the formula
(\ref{zeta}) one can, in principle, directly calculate $\zeta'(0)$ and find the function
$F(x)$ entering eq.(\ref{12}). In reality, however, it is
 technically difficult
because we know only integral representation for the heat kernel on
$AdS_2$ and integral-infinite sum form for the heat kernel on $S_2$.
Below we therefore 
 find an approximate
form of the function $F(x)$ that is enough for our purposes.

In the regime of large  or small $x$
we may find the asymptotic behavior of the function $F(x)$.
In the limit $x\rightarrow \infty$ ($r^2\rightarrow \infty$)
the heat kernel on sphere can be approximated by the heat kernel 
on  flat space $R_2$ 
which is the first term in the expansion (\ref{expansion}).
The function  $\zeta (s, r^2,l^2)$  then is approximated by
$\zeta$-function $\zeta_0 (s, r^2, l^2)$  on the product space $R_2\times H_2$.  Since we have that $\zeta_0 (s, r^2,l^2 )=
{r^2\over l^2}l^{2s}\zeta_0(s, 1,1)$  
the calculation of the effective 
action is straightforward 
and we find that
\begin{equation}
F(x)=-{1\over 60\pi}x \ln x +\zeta_0'(0,1,1)x
\label{F}
\end{equation}
for large $x$. The similar analysis for $x\rightarrow 0$ gives us that
$F(x)=O(1/x)$.

An important property of the functions $F(x)$ and $C(x)$ is how they  
transform under the inverse transformation $x\rightarrow {1\over x}$.
The analysis based on the scaling property (\ref{prop})
of $\zeta$-function and the analyticity between the heat kernels on
$S_2$ and $AdS_2$ shows that
\begin{eqnarray}
&&C({1\over x})=C(x) ~~,\nonumber \\
&&F({1\over x})=F(x)+C(x)\ln x~~,
\label{**}
\end{eqnarray}
where $x={r^2\over l^2}$.  Taking the function $C(x)$  in the form (\ref{*3})
one solves the eq.(\ref{**}) explicitly as follows
\begin{equation}
F(x)=-{1\over 60 \pi}(x-{5\over 6})\ln x+F_0(x)~~,
\label{FF}
\end{equation}
where $F_0(x)=F_0(1/x)$. It follows that $F_0'(1)=0$ and hence
$F_0(x)\sim (x-1)^2$ near $x=1$. Comparing (\ref{FF}) and (\ref{F})
we find that the leading term in $F_0(x)$ 
\begin{equation}
F_0(x)\simeq \zeta_0 '(0,1,1)(x+{1\over x})
\label{F0}
\end{equation}
is proportional  (up to an irrelevant constant)
 to the function $C(x)$. 
Thus, this term can be absorbed in the re-definition of the constant $\lambda (\mu )\rightarrow 
\lambda (\mu )-\zeta_0 '(0,1,1)$ and we may use the expression $F(x)=-{1\over 60 \pi}(x-{5\over 6}) \ln x $ as an approximation for the function $F(x)$ good
both for $x\sim  1$ and $x>>1$.

The functional (\ref{12})
can be easily  re-written as a   $2d$ covariant theory
in terms of the dilaton field $r$ and the $2d$ curvature $R$
\begin{equation}
\Gamma_0[r,R]=\int_{M^2}\left(C(r^2R)\ln {r^2\over G}+\lambda 
C(r^2R)+F(r^2R)\right)R
\sqrt{\gamma} d^2z~~.
\label{13}
\end{equation}
Note the difference of (\ref{13}) and the $2d$ effective action (\ref{6})
in the term proportional to $\ln r$. This is due to different
structure of the logarithmic UV divergences in two and four dimensions.

Now we may find the desired equations
by varying the functional (\ref{1}), (\ref{11}), (\ref{13}) with respect to $r$ and $\gamma_{\alpha\beta}$ ignoring the variation of the term $w[\nabla r, \nabla R]$
in (\ref{11}). An equivalent but simpler way is to  consider 
the functional $W_Q$ (\ref{1}) on a space $AdS_2\times S_2$ with arbitrary radiuses $l$ and $r$. Variables $l$ and $r$
then become the only gravitational degrees of freedom left. Variation with
respect to $l$ and $r$ gives us the equations we are looking for.
Following this  way we find 
\begin{equation}
W_Q[l,r]=\left( {1\over 2G}(r^2-l^2U(r))-{1\over 2}C({r^2\over l^2})
\ln {r^2\over G}-\lambda {1\over 2}C({r^2\over l^2}) -{1\over 2}F({r^2\over l^2})\right) v
~~
\label{14}
\end{equation}
for the ``quantum'' action functional.
It gives us the equations
\begin{eqnarray}
&&r^2-{l^2\over 2}rU'(r)=G\left(C(x)+xC'(x)\ln {r^2\over G}+\lambda xC'(x)+
xF'(x)\right) \nonumber \\
&&l^2 U(r)=G\left(xC'(x)\ln {r^2\over G}+\lambda xC'(x)+xF'(x)\right)~~,
\label{15}
\end{eqnarray}
where $x={r^2\over l^2}$. For the potential\footnote{
According to \cite{KS} the potential $U(r)$ in (\ref{6}) is modified by
the quantum corrections. We do not consider this possibility here
though the eqs.(\ref{15}) can be analyzed for arbitrary $U(r)$.}
 $U(r)=1-{Q^2\over r^2}$
we find that eqs.(\ref{15}) take the form
\begin{eqnarray}
&&r^2-l^2=GC(x)~~, \nonumber \\
&&r^2={Q^2\over x  }+G\left(C(x)+xC'(x)\ln {r^2\over G}+\lambda xC'(x)+
xF'(x)\right) ~~.
\label{16}
\end{eqnarray}
Analyzing these equations we first observe that when the logarithmic
UV divergences are absent in the four-dimensional theory, i.e. $C(x)\equiv 0$,
the solution of  the eqs.(\ref{16}) is especially simple
\begin{equation}
l^2=r^2=Q^2~~,
\label{17}
\end{equation}
where we took into account that in this case $F(x)=F_0(x)$ and $F_0'(1)=0$.
It is exactly the Bertotti-Robinson
solution (\ref{*}) of the classical Einstein equations. So this solution
is not deformed
by any quantum corrections if $C(x)\equiv 0$. This
observation seems to be in agreement with the non-renormalization
theorem proven in \cite{Kallosh}. Though there may appear corrections which depend strongly on
the amount of unbroken supersymmetry and correspond to a renormalization of the radius of the 
sphere, they are shown to absent for the maximally supersymmetric case.

If the function $C(x)$ is not identically zero or vanishing at $x=1$
the solution of the equations (\ref{16}) reads
\begin{equation}
{l^2\over G}={C(x)\over x-1}~~,~~{r^2\over G}={xC(x)\over x-1}~~,
\label{18}
\end{equation}
where the ratio $x={r^2\over l^2}$ should be found from the
equation
\begin{eqnarray}
&&M(x)={Q^2\over G x }~~, \nonumber \\
&&M(x)\equiv {C(x)\over (x-1)}-xC'(x)\ln ({xC(x)\over x-1})
-\lambda x C'(x)-
xF'(x)~~.
\label{19}
\end{eqnarray}
Since the eqs.(\ref{18}), (\ref{19}) are defined for $x>1$ the radius of
the spheric component is always bigger than the radius of the Anti-de Sitter
space. For $x\rightarrow 1$ the function $M(x)$ goes to infinity as follows
\begin{equation}
M(x)\simeq {1\over 180\pi}{1\over (x-1)}
\label{29}
\end{equation}
while for $x\rightarrow +\infty$ the asymptote is
\begin{equation}
M(x)\simeq {1\over 60\pi}(1+\ln 60\pi-\lambda )x+{1\over 72\pi}+O({1\over x}\ln x)~~.
\label{30}
\end{equation}
Analyzing behavior of the function $M(x)$ we find that it  depends crucially
on value of the coupling $\lambda$. For large positive $\lambda$
it  monotonically decreases and goes from $+\infty$ at $x=1$ to $-\infty$ at $x=+\infty$ and vanishes at one point $x=x_0$. The root $x_0$ is very close to $1$
and the difference $(x_0-1)$ vanishes for large $\lambda$.
At $\lambda \simeq 6.3$ the monotonic behavior of $M(x)$ changes and
for $\lambda < 6.3$ the function develops  local minimum and maximum. The local minimum becomes lower when $\lambda$ decreases
but it never hits zero. As it is seen from the asymptote (\ref{30}), at 
value $\lambda_{cr}=1+\ln 60\pi\simeq 6.239$ the behavior of the function
$M(x)$ changes drastically: it  increases linearly at large $x$ 
for $\lambda<\lambda_{cr}$. Though $M(x)$ still develops a minimum
at some point it never vanishes for  $\lambda<\lambda_{cr}$.
The form of the function $M(x)$ for different $\lambda$ is shown on 
Figure 1.

\begin{figure}
\let\picnaturalsize=N
\def\picsize{6.0in}
\def\picfilename{Planck.eps}
\ifx\nopictures Y\else{\ifx\epsfloaded Y\else\input epsf \fi
\let\epsfloaded=Y
\centerline{\ifx\picnaturalsize N\epsfxsize \picsize\fi \epsfbox{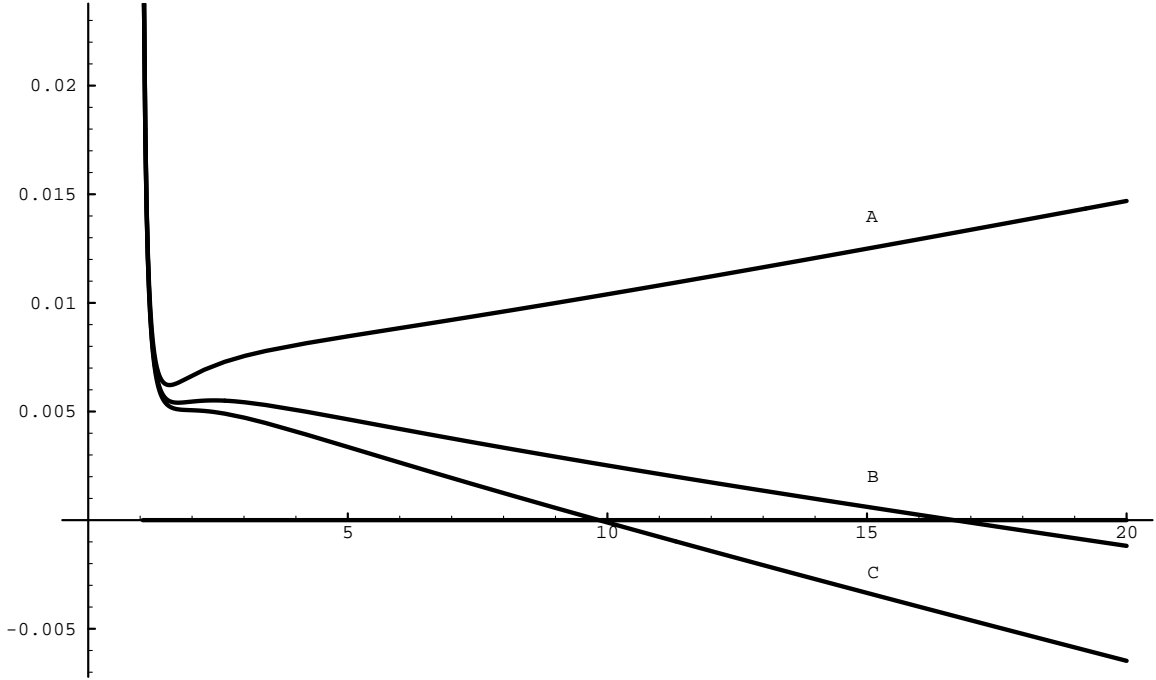}}}\fi
\caption{Plot of the function $M(x)$ for different values of
$\lambda$: (A) $\lambda=6.15$; (B) $\lambda=6.3$, (C) $\lambda=6.35$.}
\end{figure}

Roots of the equation $M(x_0)=0$ describe configurations with zero charge $Q$.
The above analysis shows that $x_0$ is close to $1$ for large positive
$\lambda$ and we find from (\ref{18}) that
\begin{eqnarray}
&&l^2\simeq {G\over 60\pi}\left({1\over 3(x_0-1)}+(x_0-1)^2\right)~~, \nonumber \\
&&r^2\simeq {G\over 60\pi}\left({1\over 3(x_0-1)}+{1\over 3}+x_0(x_0-1)\right)~~.
\label{R}
\end{eqnarray}
On the other hand,  $x_0$  becomes infinitely large when
$\lambda$ approaches the critical  value $\lambda_{cr}$. From eq.(\ref{18})
we find that in this case
\begin{equation}
l^2\simeq {G\over 60\pi}(1-{2\over 3x_0(\lambda )})~~,~~r^2\simeq {G\over 60\pi}x_0(\lambda)~~.
\label{31}
\end{equation}
Note, that there is no configuration with zero charge if $\lambda<\lambda_{cr}$.

In the charged case we are looking 
for the roots of the equation $M(x)={Q^2\over Gx}$. The analysis above shows that
for any charge $Q^2>0$ there always exists such a root $x_1\simeq 1$,
the difference $(x_1-1)$ vanishes when  charge $Q$
becomes infinitely large.
For large positive $\lambda$ there is only this solution of the equation (\ref{19}).
It is different for $\lambda<\lambda_{cr}$. Then there appears
additional root $x_2$ which grows to infinity when charge $Q$ becomes large.  Using the asymptotes (\ref{29}) and
(\ref{30}) we find in the limit of large $Q$
\begin{equation}
x_1\simeq 1+{G\over 180 \pi Q^2}
\label{32}
\end{equation}
for the first root and
\begin{equation}
x_2\simeq {Q\over \sqrt{G}\sqrt{\gamma}}~~,
\label{33}
\end{equation}
where $\gamma={1\over 60\pi}(1+\ln 60\pi-\lambda )$, for the second root.
The first root  exists for any  $\lambda$ and
describes the configuration $AdS_2\times S_2$ with radiuses
\begin{eqnarray}
&&l^2_1\simeq Q^2+{1\over 10800\pi^2}{G^2\over Q^2}~~, \nonumber \\
&&r^2_1\simeq Q^2+{G\over 180\pi}+{1\over 10800\pi^2}{G^2\over Q^2}~~
\label{34}
\end{eqnarray}
and is a quantum deformation of the classical solution (\ref{*})
analytically governed by Newton's constant $G$$ (\sim l^2_{pl})$.  The second root  (\ref{33})  appears if $\lambda < \lambda_{cr}$ and corresponds to the configuration with
the radiuses
\begin{equation}
l^2_2\simeq {G\over 60\pi}-{1\over 90\pi}{\sqrt{\gamma}G^{3/2}\over Q}~~,~~
r^2_2\simeq {G^{1/2}Q\over 60\pi\sqrt{\gamma}}~~,
\label{35}
\end{equation}
which are not analytical with respect to $G$. It is an interesting feature of the 
solution (\ref{35}) that when charge $Q$ grows the radius of the Anti-de
Sitter  component tends to the fixed  ($\sim l_{pl}$) value while the radius of the
spheric component grows  as $Q^{1/2}$.
So, for large values of $Q$ the spheric component
may have macroscopic size while
the size of the hyperbolic component remains
Planckian.

Thus, we have found that i) for any value of $\lambda$ the classical solution (\ref{*}) is modified by quantum corrections according to (\ref{34});
ii) if $\lambda>\lambda_{cr}$ there exists  also
uncharged configuration with
radiuses $r$ and $l$ determined by (\ref{R}) for large $\lambda$  or (\ref{31})
for $\lambda$ approaching $\lambda_{cr}$; iii) if $\lambda<\lambda_{cr}$ there
does not exist anymore the $Q=0$ configuration but appears new 
charged configuration  (\ref{35}) with radiuses of the Planck order. The configurations (\ref{R}),
(\ref{31}) and (\ref{35}) are absent in the classical theory and lie completely
in the Planckian domain.

The analysis we present  can be easily generalized 
for different quantum fields including gravitons.
The only information one should know  about the effective action
is how it behaves for large values of the ratio $x={r^2\over l^2}$.
If the corresponding function $M(x)$ (\ref{19})
decreases to $-\infty$ for large $x$
then we have $Q=0$ solution of the type ii). Otherwise, 
if $M(x)$ grows to $+\infty$ there exists the $Q\neq 0$ configuration of the type iii).  The solution of the type i) exists in any case.

What is the physical meaning of the Planckian $AdS_2\times S_2$
spaces we have found? In particular, what is the interpretation of the configuration with
zero electric charge? In order to answer these questions it is useful to start with the classical solution (\ref{*}). Usually, it is interpreted
as describing the geometry near horizon of the charged black hole in the extreme limit when inner and outer horizons merge. In the classical Einstein gravity the uncharged (Schwarzschild) black hole has only outer horizon. Therefore
there is no such a phenomenon as extreme black hole if  the
electric charge
vanishes. In the semiclassical
theory the situation may be different and the uncharged black hole may have
an inner horizon staying at the Planck distance from the origin.
Then the space $AdS_2\times S_2$ with zero charge $Q$  may be interpreted as describing the extreme limit of the semiclassical
Schwarzschild black hole. 
As we have seen, the $Q=0$ configuration (\ref{R}) or (\ref{31})
is  an analytical  continuation of  $Q\neq 0$ solution and should have the 
similar interpretation.
Thus, our result is an indication
that the Schwarzschild black hole has an inner Planck size horizon
in  the quantum case\footnote{Note, that the existence of the inner (with 
the size of the Planck oder) horizon in the semiclassical Schwarzschild black hole is an important ingredient  in the  scenario of the gravitational
collapse without information loss proposed in \cite{Frol-Vilk}.}.
Note, that in our consideration it is valid only for certain (though wide)
range of values of the higher curvature coupling $\lambda$.

The interpretation of the  charged solution  additional to
the (quantum deformed) Bertotti-Robinson space  is less straightforward.
Existence of this solution likely means  that there may appear different 
black hole configurations with the same value of the charge $Q$
(indicating, thus, that  Birhkoff's Theorem is not valid 
 in the semiclassical
case).  So,  for  large $Q$ there  may exist
a  near-extreme black hole configuration  with macroscopic size 
$\sim Q^{1/2}$ but Planckian curvature.  This configuration  is
absent classically and  is additional to the usual
(quantum deformed) Reissner-Nordstrom configuration
having macroscopic size proportional to $Q$ and  the curvature
invariants  bounded
by $1/Q^2$.

\bigskip

I would like to thank Oleg Zaslavskii for discussions  that stimulated
the writing of this notes and Andrei  Zelnikov for  discussing
the structure of the effective action.
I have  also benefited  from remarks by V. P. Frolov,  J. S. Dowker
and I. Avramidi.
This work was supported in part by the Natural Sciences and 
Engineering Research Council
of Canada.

\bigskip

{\bf Note added:} \\
It is interesting to note that our analysis can be extended to
describe the configurations of the type $S_2\times S_2$ (in
the Minkowskian signature it is space $dS_2\times S_2$).
One should just make the substitution $l^2\rightarrow -l^2$ ($x\rightarrow -x$),
$C(x)\rightarrow \hat{C}(x)=-C(-x)$ in the equations (\ref{16}),
the parameter $l$ now is the radius of the second sphere.
In result we find  $r^2=G{x\over x+1}\hat{C}(x)~,~~l^2=G{1\over x+1}
\hat{C}(x)$ for the radiuses where $x$ should be determined from
the equation
$$
\hat{M}(x)={Q^2\over Gx}~~,
$$
$$
\hat{M}(x)={1\over x=1}\hat{C}(x)+x\hat{C}'(x)\ln ({x\hat{C}\over x+1})+\lambda x\hat{C}'(x)+
x\hat{F}'(x)~~.
$$
For the massless field with $C(x)$ in the form (\ref{*3})
we have $\hat{C}(x)={1\over 60\pi}(x+{1\over x}+{5\over 3})$ and
$\hat{F}(x)=-{1\over 60\pi}(x+{5\over 6})\ln x+\hat{F}_0(x),~~\hat{F}_0(1/x)=\hat{F}_0(x)$.

The function $\hat{M}(x)$ is monotonic with asymptotes
$\hat{M}(x)={1\over 60\pi}(\lambda-\lambda_{cr})x+{1\over 72\pi}+O(1/x)$ for  $x\rightarrow \infty$
and $\hat{M}(x)=-{1\over 60\pi}(\lambda-\lambda_{cr})x^{-1}-{1\over 72\pi}+O(x)$ for  $x\rightarrow 0$
($\lambda_{cr}$ is the same as in the text).
It follows that for small $x$ we have solution with ${x\over 72\pi}={\lambda_{cr}-\lambda\over 60\pi}-{Q^2\over G}$ and, in particular, for $Q^2={G\over 60\pi}(\lambda_{cr}-\lambda )~~(\lambda < \lambda_{cr})$
we have $x=0$. The corresponding configuration is $R_2\times S_2$ with $l^2=\infty$ and $r^2={G\over 120\pi}$.
On the other hand, for large $x$ we have $x^2={Q^2\over G}60\pi (\lambda-\lambda_{cr}) ~~(\lambda>\lambda_{cr})$
and $r^2={x\over 60\pi}~,~l^2={G\over 60\pi}$ that is similar to the $AdS_2\times S_2$ configuration (\ref{35}).

It follows from the above asymptotes that the function $\hat{M}(x)$ should vanish at
some point. Remarkably, it happens exactly at $x=1$. This fact does not depend on
the type of the quantum massless field and is a simple consequence of the identities (\ref{**})
(indeed, we have then  $\hat{C}'(1)=0$ and $\hat{F}'(1)=-{1\over 2}\hat{C}(1)$
that results in $\hat{M}(1)=0$). The corresponding (uncharged) configuration 
is $S_2\times S_2$ with $r^2=l^2=G{1\over 2}\hat{C}(1)$.
Configurations of this type are also a subject of the consideration in \cite{ZAS}.

\end{document}